\documentclass[sn-mathphys-num,iicol,
pdflatex]{sn-jnl}

\usepackage{graphicx}%
\usepackage{amsmath,amssymb,amsfonts}%
\usepackage{natbib}
\usepackage[title]{appendix}%
\usepackage{xcolor}%
\usepackage{textcomp}%
\usepackage{booktabs}%
\usepackage{ulem, cancel,ifthen} 
\usepackage{subcaption}
\usepackage{upgreek}
\usepackage{geometry}
\newcommand{\added}[1]{{\color{red}#1}}
\newcommand{\delete}[1]{{\color{gray}\sout{#1}}}

\newcommand{\meter}{m}

\newcommand{\micro}{$\upmu$}
\newcommand{\SI}[2]{#1\,#2}

 \renewcommand{\added}[1]{#1}
\renewcommand{\delete}[1]{}

\begin{document}

\title[Article Title]{Simultaneous free-surface profilometry and subsurface velocimetry with fringe projection and PIV}


\author[1]{\fnm{Ali} \sur{Semati}}
\author[1]{\fnm{Adharsh} \sur{Shankaran}}
\author[1,2]{\fnm{Benjamin K.} \sur{Smeltzer}}
\author[1]{\fnm{Eirik} \sur{{\AE}s{\o}y}}
\author[1]{\fnm{R. Jason} \sur{Hearst}}
\author*[1]{\fnm{Simen {\AA}.} \sur{Ellingsen}} \email{simen.a.ellingsen@ntnu.no}

\affil[1]{\orgdiv{Department of Energy and Process Engineering}, \orgname{Norwegian University of Science and Technology}, \orgaddress{\street{Kolbj{\o}rn Hejes vei 2}, \city{Trondheim}, \postcode{7034} \country{Norway}}}
\affil[2]{\orgdiv{Department of Ships and Ocean Structures}, \orgname{SINTEF Ocean}, \orgaddress{Professor J.H.L. Vogts veg 1A}, \city{Trondheim}, \postcode{7052} \country{Norway}}


\abstract{
This work presents a novel combination of two well-established techniques: fringe projection profilometry (FPP) and particle image velocimetry (PIV). Despite seemingly conflicting requirements---FPP requires an opaque surface to project onto, while PIV requires a transparent fluid---both requirements are met by adding low concentrations (4--25\,mg/L) of fluorescein dye to the water. This dye strongly absorbs the blue light projected onto the surface for FPP while remaining nearly transparent to the green PIV laser, achieving simultaneous opacity and transparency depending on wavelength. A set of three optical filters suppresses fluorescence-induced noise in the PIV images and specular reflections in the profilometry images, enabling clean simultaneous acquisition. Validated against point laser-induced fluorescence measurements of the surface, the method achieves a mean absolute error in surface elevation of 18\,$\upmu$m at a dye concentration of 12\,mg/L, above which further increases in concentration yield little improvement. PIV correlation values remain robust up to 20\,mg/L. The technique resolves surface features differing in amplitude by two orders of magnitude and is demonstrated on flow past a cylinder interacting with surface waves and on droplet impacts on a quiescent surface.
}


\maketitle

\section{Introduction}\label{sec:intro}

Turbulence beneath a free water surface is of high scientific and practical importance. The water-side turbulence controls the exchange of gas and heat between ocean and atmosphere \citep{Zappa2007,Veron2011,Dasaro2014} and is widely studied numerically \citep[e.g.][]{Kermani2011,Herlina2019,Pinelli2022} and experimentally \citep[e.g.][]{Herlina2008,Turney2013,Bullee2024,Li2025,Shankaran2025}. The free surface shows distinctive, readily visible imprints of the vortex structure below, and the understanding of the intricate relationship between the two has evolved steadily from primarily qualitative observations to increasingly quantitative studies \citep[][]{Rood1995,Longuet-Higgins1996,Brocchini2001,Banerjee1994,Babiker2023,Aarnes2025}, yet there are many more questions than answers. An attractive prospect is remotely observing subsurface turbulence using the free surface as proxy \citep{Muraro2021}, since measurements from above with optical means are fast, inexpensive, and versatile compared to in-situ measurements penetrating the surface. Free-surface motions have been used to infer information about the bottom conditions of open-channel flows, such as bathymetry \citep{Dolcetti2016,Gakhar2020,Gakhar2022} and submerged canopy \citep{Mandel2017,Mandel2019}. 
    
To understand their interplay, simultaneous high-resolution measurements of the moving free surface and the velocity field underneath are therefore a highly valuable tool for studying the surface-bulk interactions experimentally. In addition to validating numerical work \citep{Shen1999,Guo2010}, such experiments also provide physical insight unconstrained by the modelling limitations of simulations. This approach allows one to study not only statistical relationships between surface motion and bulk turbulence, but also the instantaneous relationships between vortices and their signatures. However, such measurements pose a very different set of challenges compared to the more traditional measurements of subsurface turbulence in a vertical plane, where the surface can be traced with fluorescent dye \citep{Smeltzer2023,Tenhaus2024}.

Techniques for measuring the moving free surface may be broadly grouped into three classes based on their main measurement principle \citep{vanMeerkerk2020,Gomit2022}: stereoscopic, deflection, and projection methods. We briefly review these approaches and their previous combinations with velocity measurements. A fuller overview of surface measurement techniques may be found in the recent survey by \citet{Gomit2022}.

Stereoscopic methods track markers of the surface using images from at least one pair of synchronized cameras. A wide variety of markers have been used, ranging from buoyant particles \citep{Douxchamps2005}, surface ripples \citep{lePage2024}, fluorescent dye \citep{Ihrke2005} and projected patterns \citep{Tsubaki2005}, to temperature differences \citep{Hilsenstein2005,Savelyev2018} and even oranges \citep{Bjornestad2021}. Though they differ in the features they track, these methods invariably use the stereo camera system to triangulate the positions of the markers, resulting in a point cloud to which a surface can be fitted. Simultaneous measurement of surface velocity may be achieved by tracking the features between subsequent frames \citep{Douxchamps2005, Aubourg2017, Fujita2007} while subsurface velocity measurements are rather more involved. \citet{Turney2009} introduced a technique for the simultaneous measurement of surface topography and subsurface velocity for the study of microscale breaking waves. They used a stereo camera setup that imaged fluorescent particles from above the water surface. Refraction of light through the wavy interface distorted the images from the cameras. Cross-correlation of the stereo pair of images resulted in a displacement field that they related to the surface elevation from prior calibration. A second cross-correlation, performed on subsequent temporal frames, provided a velocity field; this was then corrected for refractive distortion using the surface topography calculated in the first step. In this way, the technique draws on principles characteristic of what we refer to as deflection methods.  

Ray deflection methods infer free-surface gradients from the refraction or reflection of light at the air-water interface. An early example is the colour-based technique of \citet{Zhang1994}, in which a translucent coloured screen and Fresnel lens, positioned underwater, project collimated coloured beams upwards through the water surface. The colour observed by a camera above uniquely encodes the local surface slope. Although this method is capable of measuring large surface gradients of up to 51$^\circ$e, rare among deflection techniques, the requirement that only vertical rays reach the camera, in addition to a complex calibration procedure, limits its practical applicability. \citet{Dabiri2001} later combined this approach with PIV to correlate the free-surface elevation (obtained by integration) with the subsurface velocity field in a shear layer. A similar combination of deflection measurements and PIV was implemented by \citet{Savelsberg2006}, who used a scanning laser beam refracted by the free surface to obtain line measurements of the surface gradient. By invoking Taylor's frozen turbulence hypothesis, they reconstructed two-dimensional gradient fields from these line measurements.  

Among deflection techniques, Background Oriented Schlieren (BOS) \citep{Moisy2009} is one of the most widely used, due to its simplicity, accuracy and ease of implementation. In BOS, a reference pattern is placed either above or below the free surface, while a camera observes the pattern from the opposite side, through the interface. For sufficiently small deflection angles, the apparent displacement of the pattern is linearly proportional to the surface gradient, integration of which yields the surface elevation to within an unknown offset. This difficulty in recovering the absolute surface height is a fundamental limitation shared by deflection-based methods, although it can be overcome by anchoring the reconstruction to a single point of known absolute height within the domain.

A notable variant of BOS was demonstrated by \citet{Fouras2008} and later \citet{Gomit2013}, who showed that PIV seeding particles can themselves serve as the reference pattern when the laser sheet is oriented parallel to the free surface. A camera below the channel captures standard double-frame PIV images, while a camera above the surface captures a single frame synchronised with one of the two PIV frames. Cross-correlating the simultaneously captured frames from the two cameras yields the surface gradient, obtained at almost no additional cost from an otherwise standard PIV setup, though spatial resolution is limited by the requirement for sufficient seeding density within each interrogation window.

A more severe limitation of deflection methods arises when surface slopes become too large. This causes ray crossing (caustics), which destroys the unique mapping between the pattern and its image and prevents accurate estimation of the displacement. Even in the absence of ray crossing, strong distortions can cause displacement estimation algorithms to fail. As a result, many flows of interest, such as wind-generated surface ripples, are outside the practical range of deflection methods. 

Projection-based techniques can accommodate much larger surface gradients. A widely used approach illuminates the free surface with a thin laser sheet to obtain profile measurements. \citet{Bonmarin1989} used this technique for qualitative measurements of breaking waves, and \citet{Duncan1999} added fluorescein dye to the water, which enabled automatic profile detection. This approach, now commonly referred to as laser-induced fluorescence (LIF), is readily combined with PIV when measurements are made in a vertical plane \citep{Buckley2017}. In this configuration, the same laser can be used for both PIV and LIF, with optical filters preventing cross-contamination between cameras. More recently, \citet{vanMeerkerk2020} extended LIF by scanning the laser line in the transverse direction, producing three-dimensional point clouds of the free surface, though with limited resolution in the scanning direction.

Structured light methods analyse the distortion of a pattern projected onto the surface, relative to the same pattern on a flat reference plane. While the projected pattern may take different forms, the underlying principle remains the same: when the incident light is not normal to the surface, a change in surface height induces a lateral shift in the point where the ray intersects the surface, which is recorded by a camera and analysed algorithmically. Various patterns have been employed, including dots, lines, grids and sinusoidal fringes, with a trade-off between robustness and spatial resolution. Simpler patterns, such as dot grids, are computationally efficient and robust but limited in resolution by the dot spacing, whereas sinusoidal fringe patterns provide per-pixel height measurements at the cost of increased computational complexity and sensitivity to large gradients.

The development of Fourier transform profilometry by \citet{Takeda1982} enabled efficient automatic analysis of fringe distortions and has since been applied to a wide range of free-surface flows, including vortex-induced surface depressions \citep{Zhang2002}, dam-break flows \citep{Cochard2008}, surface waves \citep{Cobelli2009}, wave turbulence \citep{Herbert2010, Cobelli2011} and liquid sprays \citep{Roth2020}. 

A key challenge for structured light techniques is the optical transparency of water, which is commonly addressed by adding titanium dioxide particles to render the fluid opaque. Titanium dioxide, however, can significantly alter surface tension \citep{Przadka2012}, and hinders all optical access to the interior of the fluid. Alternative approaches have also been explored; for example, \citet{Roth2020} employed high concentrations of fluorescein in their experiments (10\,g/L, about a thousand times higher than in the present work).

To date, structured-light projection methods have not been combined with PIV, primarily because the former require an optically opaque surface, whereas the latter requires transparency. In the present study, we bridge this gap with low concentrations of fluorescein dye (4 to 25\,mg/L) and validate the method with independent LIF measurements. The remainder of this paper is structured as follows: Section~\ref{sec:Principles} reviews the fundamentals of Fourier transform profilometry and discusses the optical properties of fluorescein dye. The experimental setup and calibration procedure are described in Sec.~\ref{sec:ExpSetup}, followed by an error analysis for varying dye concentrations in Sec.~\ref{sec:ErrorAnalysis}. After measuring the attenuation length of light in the dye solution in Sec.~\ref{sec:AttenuationLength}, we apply the method to wave-vortex interactions behind a cylinder and to droplet impacts in Sec.~\ref{sec:Applications}. Finally, we provide recommendations for implementing the technique and outline some remaining challenges in Sec.~\ref{sec:Challenges}.


\section{Principles}\label{sec:Principles}

\subsection{Fringe projection profilometry}\label{sec:FringeProjection}

Surface topography is measured by projecting a fringe pattern onto the surface and comparing its distortion to that of the pattern on a flat reference plane which defines the vertical origin. When the projected light is not normal to the surface, changes in elevation produce shifts in the fringe pattern, which are recorded by a camera above. Each observed shift is related to the local surface elevation through a pixel-wise calibration procedure. Although the specific features of the pattern do not affect this relationship, they influence the achievable measurement accuracy for a given computational cost. Sinusoidal fringe patterns are especially well suited for demodulation via the Fourier transform method of \citet{Takeda1983}. When projected onto a flat reference plane, the intensity profile is expressed as
\begin{equation}
I_\mathrm{ref} = I_0(x,y) + R(x,y)\cos(2 \pi f_0 x + \phi_0),
\end{equation}
where $I_0(x,y)$ is the background illumination intensity, $R(x,y)$ accounts for local variations in fringe amplitude, $f_0$ is the frequency of the fringe pattern, and $\phi_0$ is the initial phase offset. 

A perturbation in the surface modifies the phase, resulting in the perturbed intensity profile
\begin{equation}
I = I_0(x,y) + R(x,y)\cos(2 \pi f_0 x + \phi).
\end{equation}
The surface elevation field is encoded in the phase difference $\Delta \phi(x,y) = \phi - \phi_0$, which is recovered by computing the phase of the complex product
\begin{equation}
\Delta \phi(x,y) =  \mathrm{Im}\left[\log\left(\hat{G}_{\mathrm{ref}}^* \cdot {\hat{G}}\right)\right],
\end{equation}
where $\hat{G}_{\mathrm{ref}}$ and $\hat{G}$ are the complex signals obtained by Fourier transforming the reference and perturbed images, isolating the spectrum around the carrier frequency $f_0$ (e.g. using a Gaussian or rectangular window), and then taking the inverse transform.

The spectrum of the perturbed image exhibits sidebands around the carrier frequency, whose width increases with the maximum surface slope. For accurate demodulation, these sidebands must not overlap with either the DC component or the second harmonic. The latter condition is generally more restrictive, and \citet{Takeda1983} showed that for their specific imaging geometry, it requires 
\begin{equation}
\left| \frac{\partial h}{\partial x} \right|_{\max} < \frac{1}{3} \cdot \left( \frac{L}{D} \right),
\end{equation}
where $h$ is surface elevation, $L$ is the distance from the projector exit pupil to the reference plane, and $D$ is the distance between the projector and camera. 

The complex exponential satisfies \[e^{x + i(2n \pi + y)} = e^x e^{iy}, \quad n \in \mathbb{Z},\] so all numbers differing by integer multiples of $2\pi i$ in their imaginary part map to the same complex value. Consequently, the complex logarithm is multivalued, and in practice is restricted to its principal branch with imaginary range $(-\pi, \pi]$. When the true phase shift exceeds this range, $\Delta \phi(x,y)$ is discontinuous at the branch cut with jumps of $\pm 2\pi$, a condition known as phase wrapping.  

The simplest unwrapping method proceeds line by line through the phase field, adding or subtracting multiples of $2\pi$ whenever a jump greater than $\pi$ is encountered, so that adjacent values differ by less than $\pi$. While this row-by-row (or column-by-column) approach is computationally efficient, it is sensitive to noise and sharp phase variations. More robust techniques have been developed at the cost of additional computational complexity \citep{Herraez2002}. For our experiments, the high signal-to-noise ratio and smooth surface meant that the simple line-by-line method proved robust.  

Unwrapping produces a continuous phase field, but the result remains indeterminate up to an unknown multiple of $2\pi$. To resolve this ambiguity, the phase must be known at a minimum of one point in the domain. This can be addressed by spatially limiting the projected fringe pattern so that it does not fill the entire field of view. The resulting dark margin along one edge, parallel to the fringes, makes the first fringe line clearly identifiable. The integer multiple of $2\pi$ required to correct the phase field is then determined by comparing the position of the first fringe in the deformed image to the reference image.

The phase shift can be related to surface elevation by geometric methods, which require careful placement of the projector and camera and measurement of the distances involved \citep{Takeda1983, Zappa2009}, or polynomial methods, where images of the fringe pattern projected onto several parallel planes at known heights are used to relate phase shift to surface elevation for each individual pixel. There are advantages and disadvantages associated with each family of methods. The restrictions imposed by geometric techniques on the alignment of the projector and camera can be difficult to achieve with sufficient accuracy. For example, the method of \citet{Takeda1983} requires that the camera entrance pupil and the projector exit pupil be at the same height with respect to the surface, which is not straightforward to achieve in practice, owing partly to the fact that their exact positions are not always readily apparent. That said, if the facility allows placing the projector and camera far enough from the surface, good accuracy can be achieved with minimal effort. 

Polynomial methods for surface reconstruction from phase shift, on the other hand, avoid the strict restrictions on placement of camera and projector, but require images of the fringe pattern on several planes at known heights. Further, the calibration is valid only over the area covered by the calibration plate. In this study we used a second-order polynomial calibration, discussed further in Sec.~\ref{sec:Calibration}. For a review of calibration methods in FPP, see \citet{Feng2021}.

The profilometry results presented here were processed using custom MATLAB code (available online). The code implements both Fourier transform and wavelet transform profilometry \citep{Zhong2004, Gdeisat2006}, and supports two phase-to-height conversion methods: the equation of \citet{Takeda1983} and the polynomial method described above, for which a calibration module is included. Phase unwrapping uses the 1D line-by-line method by default, switching automatically to the 2D method of \citet{Herraez2002} if unwrapping errors are detected.

\subsection{Fluorescent dye}\label{sec:Fluorescein}

\begin{figure*}[tb]
  \centering
  \includegraphics[width=0.7\textwidth]{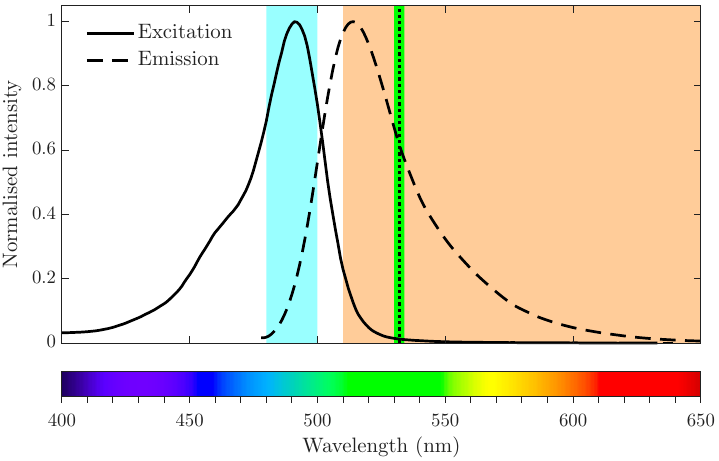}
  \caption{Excitation and emission spectra of fluorescein overlaid with the spectra of the bandpass filters used in this study. Light from the projector is filtered with a $20$\,nm wide filter at $490$\,nm (cyan). A 510\,nm long-pass filter (orange) is placed on the profilometry camera lens and fluorescent noise is reduced with a $4$\,nm wide bandpass filter (green) centred at $532$\,nm in front of the PIV camera.}
  \label{fig:fluoresceinSpectra}
\end{figure*}

The primary challenge of combining projection-based surface measurement with PIV lies in their seemingly conflicting requirements: surface measurement requires an optically opaque projection surface, whereas PIV requires a transparent medium. These requirements can be reconciled by adding a fluorescent dye to the water that absorbs the projected pattern light while remaining transparent to the laser light.  

In most facilities, including ours, PIV is performed using Nd-doped lasers (such as Nd:YAG or Nd:YLF) emitting green light at wavelengths between 527\,nm and 532\,nm. Unlike LED-based sources, these lasers provide the high power and short pulse duration (on the order of nanoseconds) required for high-accuracy PIV. We therefore restrict our choice of dye to those that interact minimally with green light at $532$\,nm. The ideal dye has the following properties:
\begin{enumerate}[\hspace{4pt}(PT)]
        \item Excitation and emission spectra do not overlap with laser light at $532$\,nm.
        \item High quantum yield. 
        \item Non-toxic and not harmful to the environment. 
        \item Readily available and cost-effective for large facilities.
\end{enumerate}
Although a wide variety of fluorescent dyes with diverse optical properties are available, most of these are produced for use in the biomedical sciences and are typically supplied in amounts on the order of milligrammes. In contrast, large-scale hydrodynamic facilities have volumes on the order of 10--100\,tonnes, requiring hundreds of grammes of dye. Condition \#4 thus limits the choice of dye, to our knowledge, to the rhodamine family or to fluorescein. Rhodamine dyes strongly absorb green light and would completely block the $532$\,nm laser light, leaving fluorescein as the only viable option.  

Fluorescein is comparatively inexpensive, produced in large quantities, biodegradable, non-toxic (allowing easy handling and convenient disposal) \citep{hara1998}, and has a high quantum yield (95\%). Its absorption and emission spectra, shown in Fig.~\ref{fig:fluoresceinSpectra}, are, however, not ideal: fluorescein absorbs light at $532$\,nm---weakly, but not insignificantly---and emits primarily in the green, $500$ to $550$\,nm range. This introduces several experimental challenges. Absorption of laser light by the dye reduces the signal-to-noise ratio (SNR) of the PIV images and limits the maximum usable dye concentration. At low dye concentrations, however, the light projected from above penetrates deeper into the water, reducing the contrast of the profilometry images and degrading the accuracy of surface reconstruction with FPP. A balance must therefore be struck between the requirements of the two techniques. Furthermore, excitation of the dye by the laser results in green fluorescence, which appears as noise in the PIV images, further reducing the SNR. We address these challenges through the addition of optical filters to the projector and both cameras, as detailed in Sec.~\ref{sec:opticalFilters}.

\subsection{FPP and semi-transparency}\label{sec:Semitransparency}
When the water surface is not opaque, light is absorbed and emitted by a subsurface volume of water instead of just the surface layer. Drawing on studies of the error caused by translucency in solids \citep{Lutzke2011, Xu2019}, we recognise two types of error introduced by semi-transparency. Both share the same physical origin, that is, light interacting with a subsurface volume rather than the surface alone, but manifest differently in the reconstructed surface. 

First, subsurface emissions reduce the contrast of the projected pattern. As light emission originates from a subsurface volume rather than the surface alone, a point source projected vertically down onto the surface appears blurred when viewed from above. In the same way, a sine-wave pattern projected onto the surface loses contrast. This loss of contrast makes the measurement system more sensitive to random noise \citep{Xu2019}, since a fixed level of random image noise becomes more significant when the contrast is reduced. Taken to the extreme, the sine wave becomes a flat line and all phase information is lost. 

\added{Second, any light ray entering the water excites a column of dye whose centroid of emission lies beneath the surface entry point. This displacement is seen as a false phase shift by the camera whenever its viewpoint differs from that of the projector.} The magnitude of the resulting phase error depends on the viewing angle, as shown by \citet{Lutzke2011}, the shape of the surface (due to refraction) and, in our case, the dye concentration. 

Both effects are governed by how far light travels into the water column before being attenuated, with the former being a limit on precision and the latter being a limit on accuracy. We analyse the effect of dye concentration on reconstruction error in Sec.~\ref{sec:ErrorAnalysis} and measure the attenuation length in Sec.~\ref{sec:AttenuationLength}. 


\section{Experimental Setup}\label{sec:ExpSetup}

\begin{figure*}[!htb]
  \centering
  \includegraphics[width=12.8cm]{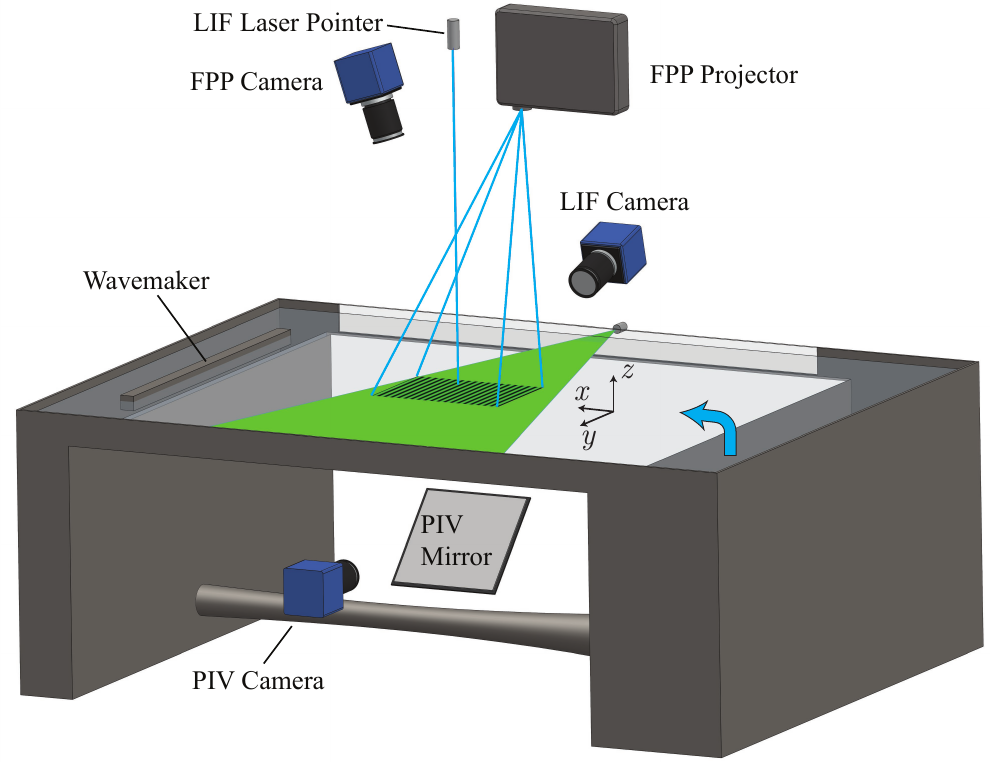}
  \caption{Schematic of the experimental setup. Bandpass filters centred at 532 and 490~nm are placed on the PIV camera and projector, respectively, and a long-pass filter at 510~nm is placed on the FPP camera. The laser pointer and LIF camera were included only in the validation experiments.}
  \label{fig:experimentalSetup}
\end{figure*}

To validate and demonstrate the measurement technique, experiments were conducted in the small recirculating water channel facility at NTNU. This facility has a $2\times2\times 0.13$\,m test section with optical access through the base and one sidewall, and is equipped with a paddle-type wave generator. A schematic of the setup is shown in Fig.~\ref{fig:experimentalSetup} and further details about the flume may be found in \citet{Smeltzer2019,Smeltzer2019a}.

Fluorescein disodium salt (Thermo Fisher Scientific) was dissolved in the water at concentrations between 4 and 25\,mg/L. All imaging was performed using LaVision CX2-25MP cameras. Illumination for PIV was provided by a double-pulsed Nd:YAG laser (Litron Nano L) with a pulse energy of 200\,mJ at 532\,nm, while sinusoidal stripe patterns (hereafter referred to as ``fringes'') were generated by a video projector (Epson EH-TW6700) with a native resolution of $1920 \times 1080$ pixels. This projector uses a high-pressure mercury vapour lamp as its light source. Although illumination at 490\,nm would be ideal, corresponding to the peak absorption wavelength of fluorescein, the projector spectrum provides sufficient intensity near this wavelength. Mercury vapour lamps have spectral peaks at 403, 435, 546 and 578\,nm, but exhibit considerable spectral broadening at higher pressures \citep{Derra2005}, enabling effective excitation of the dye.

Two sets of experiments were conducted: the first to assess the effect of dye concentration on the accuracy of surface measurements and the second to assess its effect on PIV measurements. 

To validate the FPP measurements, an independent measurement of surface elevation was obtained by projecting a single beam from a 488\,nm continuous-wave laser onto the surface. A camera fitted with a 200\,mm lens (the LIF camera in Fig.~\ref{fig:experimentalSetup}) recorded the laser spot. The high magnification allowed us to extract the surface profile across the spot width (approximately 1.5\,mm), from which the local surface elevation and slope were recovered by image analysis. This method is robust to variations in dye concentration and is detailed in Sec.~\ref{sec:pointLIF}. For each concentration, the wavemaker was operated at 1\,Hz and surface waves were measured simultaneously by both systems at 40\,Hz for 20 seconds. Each measurement was performed twice to ensure repeatability. The exposure times were set to 5\,ms for the FPP camera and 2\,ms for the LIF camera.    

The simplest configuration for simultaneous measurements aligns the PIV and profilometry cameras symmetrically with respect to the water surface, with both optical axes normal to it, but in this arrangement the bandpass filter placed on the PIV camera reflects fluorescein emissions back towards the profilometry camera (see Sec.~\ref{sec:opticalFilters} for details). This reflection appears as a circular artefact in the profilometry images and, because the light is refracted by the undulating water surface before reaching the camera, is difficult to remove through post-processing. 

This artefact can be eliminated by placing the PIV camera outside the viewing angle of the profilometry camera. For planar PIV, the optical axis must be kept normal to the laser sheet to minimise out-of-plane errors. We thus maintained the PIV camera in its normal orientation and tilted the profilometry camera by 12.5$^\circ$ to exclude the PIV camera from its field of view. The profilometry camera was placed 1\,m above the water surface. The minimum required tilt angle decreases with increasing camera separation, making it preferable to position the PIV camera as far as possible from the profilometry camera and to use a longer focal length lens to maintain the same field of view. Due to space constraints below the channel, the optical path length between the PIV camera and the laser sheet was extended to 1.7\,m using a mirror, as shown in Fig.~\ref{fig:experimentalSetup}. The PIV and profilometry cameras were equipped with $100$\,mm and $60$\,mm lenses, respectively. Note that this geometric constraint is specific to planar PIV in the horizontal plane and does not apply to PIV in the vertical plane or stereo PIV setups, where the profilometry camera can be oriented normal to the water surface. 

\added{
The PIV and FPP cameras had fields of view of $207 \times 235\, \mathrm{mm}$ and $213 \times 234\, \mathrm{mm}$, respectively. The overlap between the two measurement domains was $175 \times 220\, \mathrm{mm}$, where the reduced extent in $x$ results from an unilluminated margin in the FPP images (see Sec.~\ref{sec:FringeProjection}).
}

The effect of fluorescein dye on PIV measurements was assessed by acquiring $500$ independent PIV snapshots of uniform flow in a horizontal plane 2\,cm below the water surface for each dye concentration. Polystyrene particles with a mean diameter of $40$\,{\textmu}m were used as tracers. After applying a min-max pre-processing filter \citep{Adrian2011}, the images were processed using LaVision DaVis 11.0 with a $48 \times 48$\,pixel (2.1 $\times\ 2.1$\,mm) window and $50\%$ overlap. The correlation coefficients were spatially and temporally averaged to yield a single metric representing PIV quality.


\subsection{Optical filters}\label{sec:opticalFilters}

\begin{figure*}[tb]
    \centering
    \includegraphics[width=0.9\textwidth]{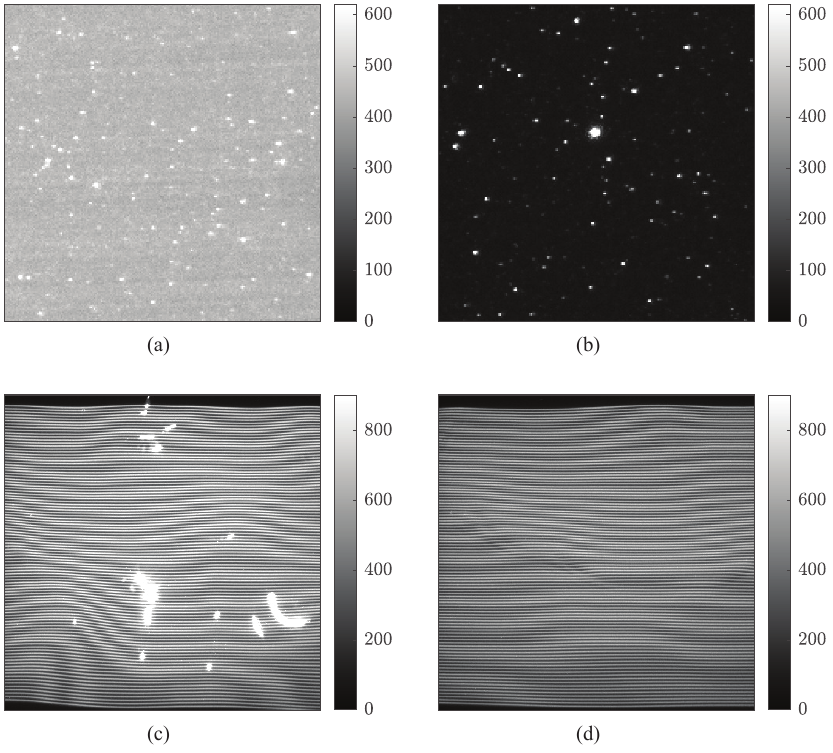}
     \caption{Optical challenges associated with this experimental approach. (a)~PIV image showing noise that stems from the absorption of laser light at 532\,nm by fluorescein. (b)~PIV image taken under the same conditions as (a), but with an ultranarrow bandpass filter, which removes most of the noise. (c)~Profilometry image of the fringe pattern showing specular reflections from the water surface, which saturate parts of the camera sensor. (d) Profilometry image with the addition of a 510\,nm long-pass filter to the camera, eliminating the specular reflections.}
    \label{fig:filterEffects}
\end{figure*}

The spectral properties of fluorescein dye present challenges for both PIV and profilometry. Figure~\ref{fig:filterEffects}a shows a PIV snapshot acquired without additional filtering. Because fluorescein has small but non-zero absorption at $532$\,nm as Fig.~\ref{fig:fluoresceinSpectra} shows, excitation of the dye by the intense laser light appears as background noise, reducing the signal-to-noise ratio (SNR). Both the laser light and the fluorescence are green, hence standard PIV bandpass filters are ineffective. However, we can exploit the difference in spectral width between the two sources: the laser light is narrowband ($532 \pm 0.25$\,nm), whereas the dye emission has a comparatively broad spectrum spanning the whole green range and beyond (see Fig.\ \ref{fig:fluoresceinSpectra}), only a tiny fraction of which lies within the laser's spectral band. We therefore employ a $4$\,nm-wide bandpass filter (RET532/4x, Chroma Technology) centred at $532$\,nm for PIV. As shown in Fig.~\ref{fig:filterEffects}b, this filter eliminates nearly all of the background noise. 

\begin{figure*}[tb]
    \centering
    \begin{subfigure}[t]{0.45\textwidth}
        \centering
        \includegraphics[width=\linewidth]{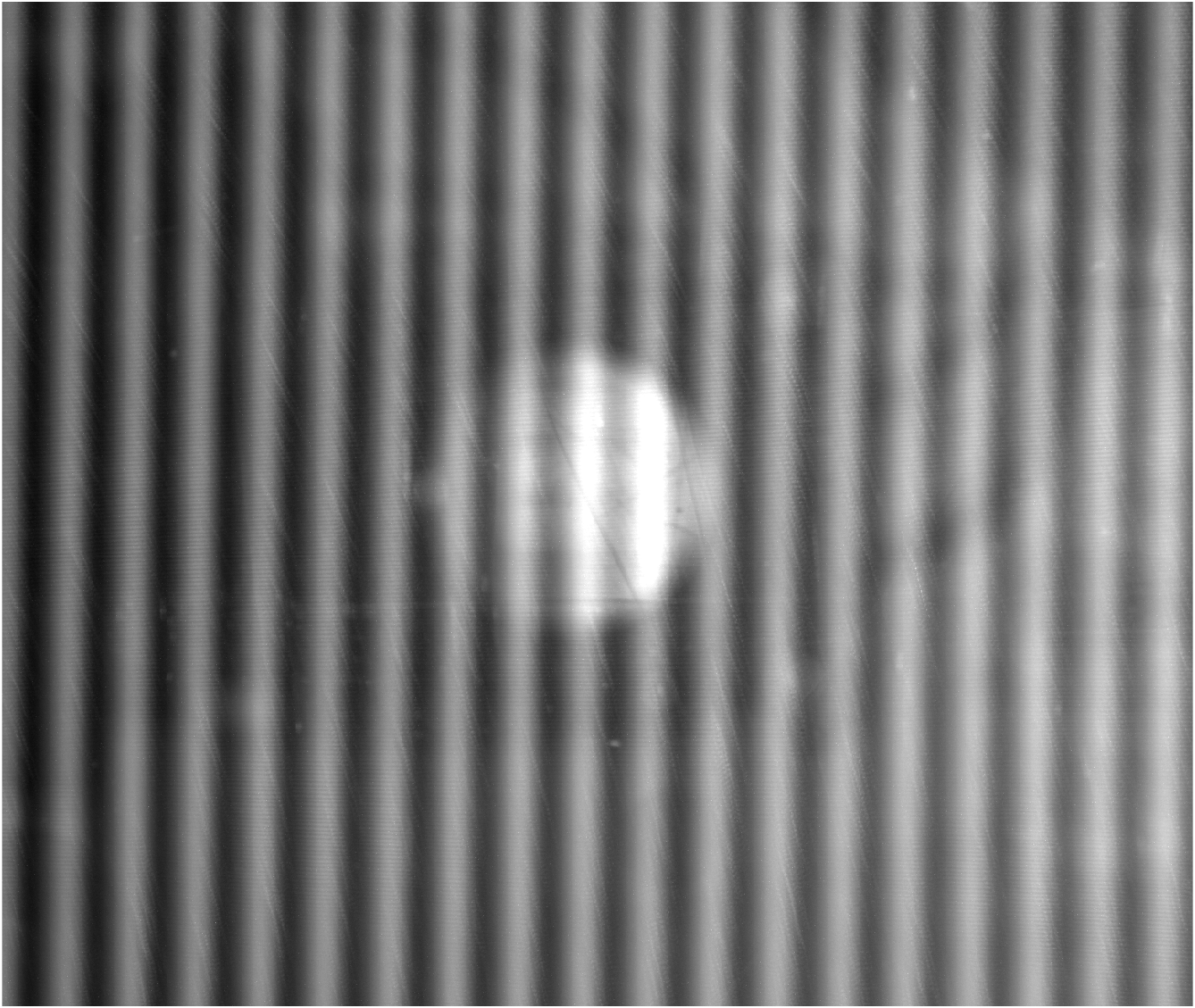}
        \caption{}
        \label{fig:ultranarrowReflection}
    \end{subfigure}%
  \hfil
    \begin{subfigure}[t]{0.45\textwidth}
        \centering
        \includegraphics[width=\linewidth]{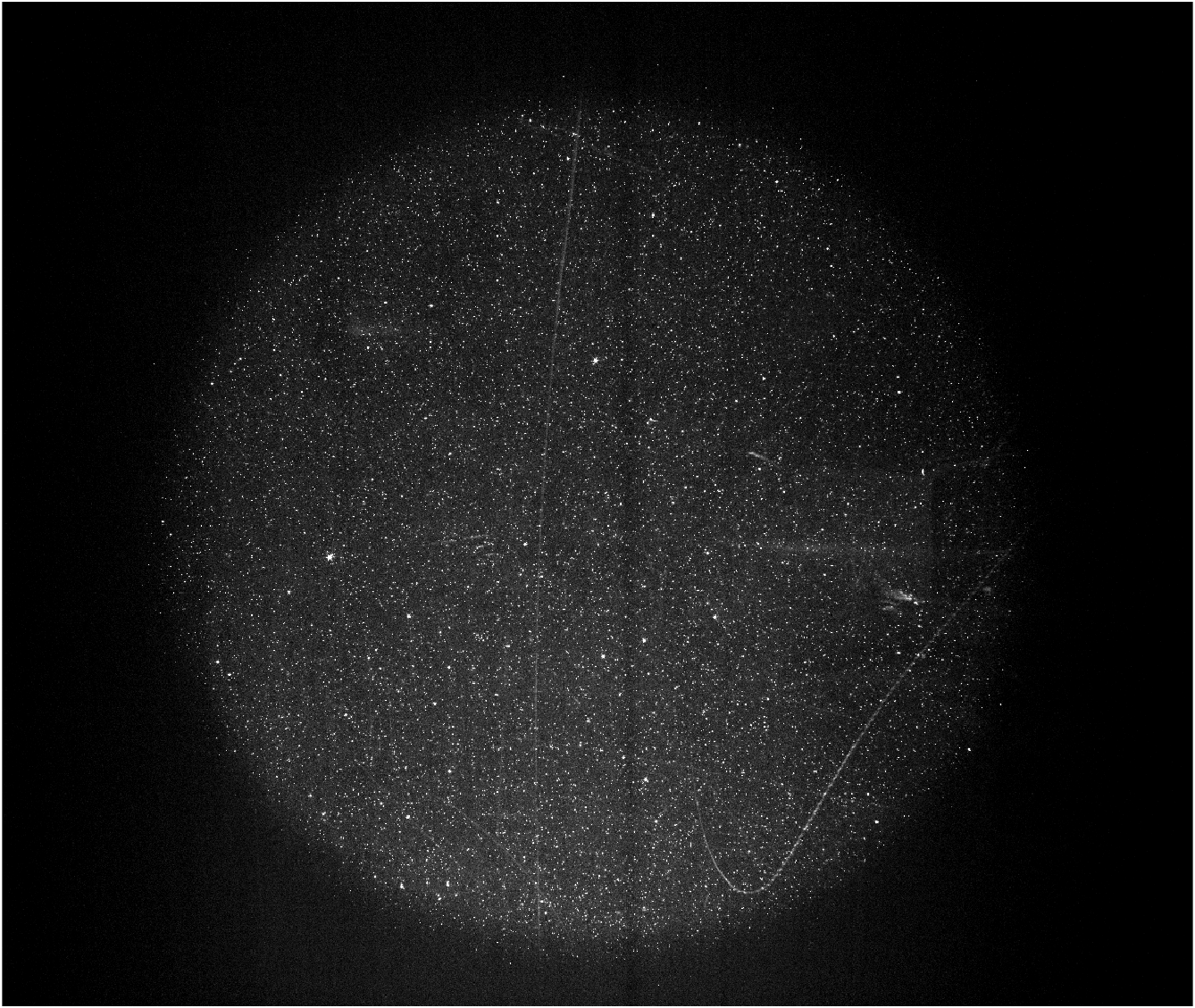}
        \caption{}
        \label{fig:ultranarrowCutoff}
    \end{subfigure}
    \caption{(a) Profilometry image taken from above the water surface showing reflection of light from the ultranarrow bandpass filter on the PIV camera below the surface. (b) PIV image showing cut-off of light when using a wide-angle lens with the ultranarrow filter.}
\end{figure*}

The narrowband filter also prevents contamination of the PIV images by the profilometry fringe pattern, which would otherwise necessitate strobing the projector in sync with PIV acquisition or using a mechanical shutter. During the short exposure time of the first frame in double-frame imaging (typically microseconds), the contribution of the projector is negligible compared to the high-intensity Nd:YAG laser pulse (nanosecond duration). However, the second frame presents a specific challenge. The exposure time of the second frame is dictated by the sensor readout time, approximately $100$\,ms for our cameras. Over this longer duration, the integrated intensity of the continuous projector light becomes significant, making the fringe pattern visible. By blocking most of the fluorescence, the narrowband filter makes the pattern invisible to the PIV camera, allowing for continuous projection and simplifying the experimental setup.

Using an ultranarrow filter does introduce two limitations. First, as noted in Sec.~\ref{sec:ExpSetup}, the filter reflects fluorescence and appears as a bright artefact to the profilometry camera unless the setup avoids mirror symmetry. Figure~\ref{fig:ultranarrowReflection} shows a profilometry image acquired with the cameras in mirror opposition, illustrating this reflection. Second, the spectral properties of the filter depend on the angle of incidence. Manufacturers typically report performance for orthogonal incidence; at oblique angles, the passband shifts, potentially blocking the monochromatic laser light. This may be encountered when using wide-angle lenses, as shown in Fig.~\ref{fig:ultranarrowCutoff}, where an image taken with a 28\,mm focal length lens exhibits a vignetting effect. The darkening on the edges is due to spectral cut-off, not physical obstruction. For PIV photography, the filter thus limits the maximum angular field of view that can be achieved.

As discussed in Sec.~\ref{sec:Fluorescein}, increasing the concentration of dye makes the surface more opaque to the projected fringe pattern, improving surface reconstruction with FPP. However, absorption of laser light by fluorescein places an upper limit on dye concentration. This limit depends on laser power, PIV particle size (larger particles scatter more light), and the path length of laser light in water (from the laser to the imaging region and from the particles to the camera). Placing a narrow bandpass filter on the PIV camera as described earlier extends the maximum dye concentration with which PIV measurements of acceptable quality can be obtained. Similarly, the minimum dye concentration for reliable profilometry can be lowered by placing a bandpass filter on the projector centred at the maximum absorption wavelength of fluorescein ($\approx 490$\,nm, Fig.~\ref{fig:fluoresceinSpectra}). At the same dye concentration, restricting the spectral composition of the projected light to those wavelengths most strongly absorbed by fluorescein will result in a lower penetration depth for the light and thus increase the contrast. We selected a bandpass filter (MV490/20, Chroma Technology) centred at 490\,nm with a full-width at half-maximum (FWHM) of 20\,nm. 

The final challenge arises from specular reflections from the water surface. These reflections, shown in Fig.~\ref{fig:filterEffects}c, can saturate the image sensor and prevent reconstruction of affected regions. Specular reflections are a common obstacle in FPP measurements of polished or metallic surfaces, and various mitigation strategies have been proposed in the literature \citep{Nayar2012, Song2017}. A common approach is to linearly polarise the projected light \citep{Salahieh2014, Dave2022}. Since specular reflections largely retain the incident polarisation, they can be filtered by a cross-oriented polariser on the camera. As the dye emits unpolarised light, this approach is applicable to our system but comes with a severe penalty: the combination of polarisers attenuates the fluorescence signal by a theoretical minimum of 75\%.  

We therefore adopt an alternative approach \citep{Roth2020}, exploiting the Stokes shift of fluorescein. Because the emission peak occurs at a longer wavelength than the absorption peak, a long-pass filter (ET510lp, Chroma Technology) at 510\,nm transmits the majority of emitted light while blocking the reflected excitation light. We observed a $23$\% reduction in signal intensity after placing the long-pass filter on the profilometry camera, consistent with the $25$\% reduction predicted by numerically integrating the emission spectrum above $510$\,nm. This long-pass filter does not, however, block the PIV laser light, and therefore the FPP camera’s exposure should not overlap with either laser pulse. This can be achieved by synchronising the FPP camera so that exposure occurs before or after the two laser pulses, or between them if the exposure time is sufficiently short.

The optical density (OD) of both the profilometry and projection filters is critical in blocking specular reflections. We found that the OD3 rating of the MV490 filter mentioned above was insufficient to prevent out-of-band transmission from the projector, which appeared as specular reflections. The ET480/40x however, centred at 480\,nm with a FWHM of 40\,nm and rated OD6, combined with the camera's OD6 long-pass filter, completely eliminated the specular reflections, as shown in Fig.~\ref{fig:filterEffects}d. To achieve both the necessary spectral narrowness and high blocking efficiency, we stacked the two projector filters. In this configuration, the ET480/40x provided the optical density required to eliminate the specular reflections, while the MV490/20 constrained the bandwidth. 


\subsection{Point LIF validation}\label{sec:pointLIF}

To validate the FPP method, we performed simultaneous, independent point measurements of the free surface using an optical wave gauge (point-LIF). A 488\,nm continuous-wave laser pointer from Zeus lasers (the LIF laser pointer in Fig.~\ref{fig:experimentalSetup}) was positioned above the water surface and oriented vertically downwards to project a spot of light onto the interface, which was imaged at $40$\,Hz. Although dye concentration affects the penetration depth of light into the water, the point-LIF method relies only on detecting a sharp intensity transition at the interface and is therefore not sensitive to this effect.

\begin{figure}[tb]
  \centering
  \includegraphics[width=\linewidth]{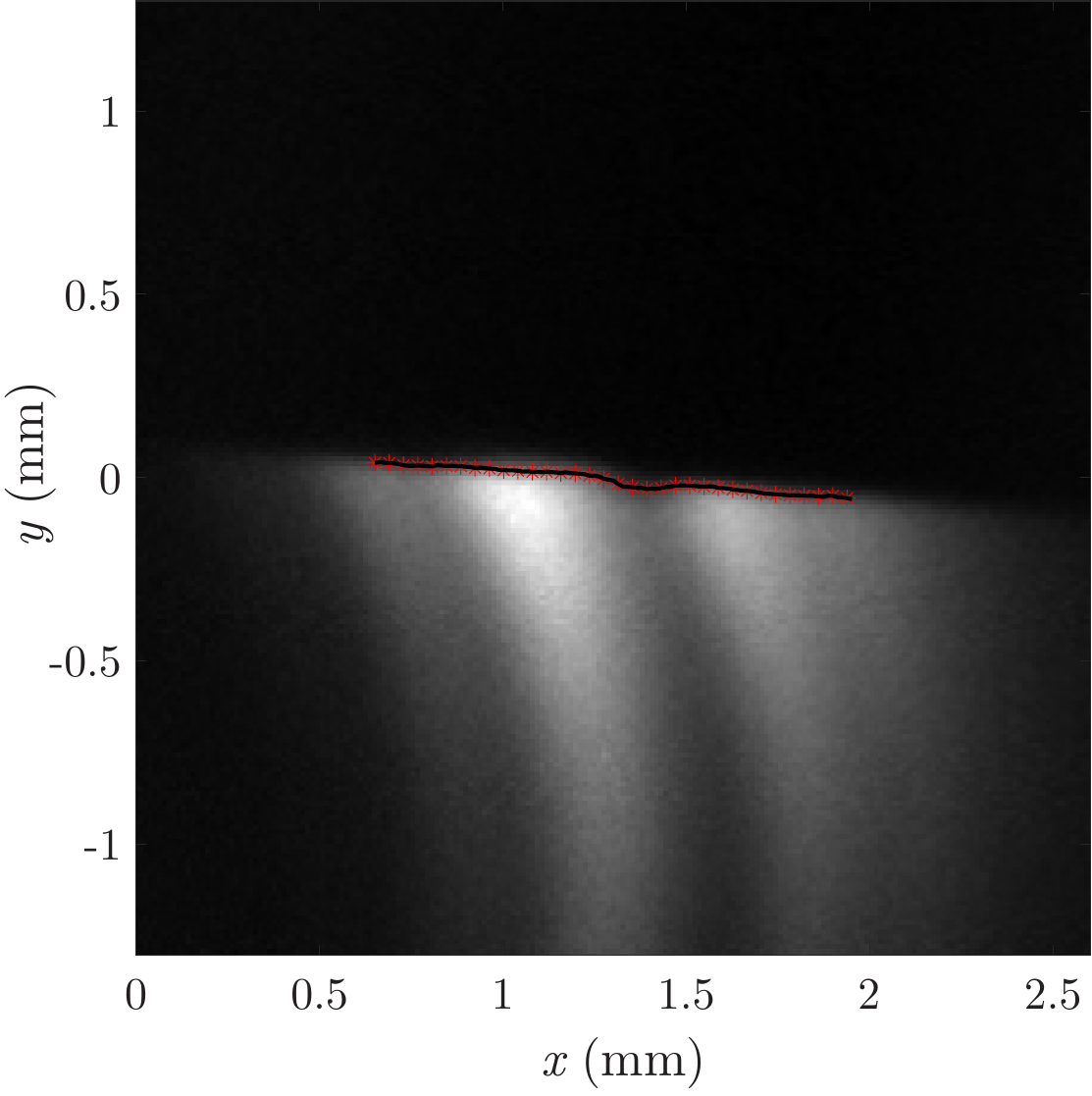}
  \caption{Representative side-view image from the point-LIF camera showing the air-water interface. A vertical laser beam illuminates the water column from above, exciting the fluorescent dye along its path. The detected air-water interface is overlaid as a black line with red markers.}
  \label{fig:pointLIF}
\end{figure}

A sharp increase in pixel intensity occurs at the air-water interface as the laser beam excites the fluorescent dye, and a simple threshold operation followed by contour detection can be used to identify the interface. But since dye concentration varies between experiments, such an approach would require case-specific thresholds. Further, the laser beam was spatially inhomogeneous, likely due to imperfect collimation or lens aberrations, as visible in Fig.~\ref{fig:pointLIF}.

We therefore opted for a gradient-based approach, commonly used for edge detection in computer vision, in which the interface was identified as the location of the maximum vertical intensity gradient. Intensity profiles were extracted along vertical columns of each image and were smoothed using a third-order Savitzky–Golay filter to reduce the noise amplification inherent in numerical differentiation. In Fig.~\ref{fig:pointLIF}, the detected free surface has been overlaid on the image with red markers. The high camera resolution (\SI{13}{\micro\meter} per pixel) provided over 100 points across the 1.5\,mm width of the laser spot, sufficient for a robust linear fit to extract both local slope and elevation. 

The high-intensity laser spot was also visible in the FPP images, preventing direct comparison at the exact measurement point. We therefore sampled the FPP data at points 4\,mm on either side of the laser spot and interpolated the value at the centre. This approach proved robust and accurate, with significant deviations observed only towards the end of each run, when wave reflections from the sidewalls degraded the interpolation accuracy (discussed further in Sec.~\ref{sec:ErrorAnalysis}).


\subsection{Calibration}\label{sec:Calibration}

The PIV and FPP cameras were calibrated using a two-level, double-sided calibration plate (LaVision Type 20) mounted on a translation stage. The bottom surface of the plate was first aligned with the laser sheet ($60$\,mm above the channel bed), after which the plate was raised to align the top surface with the quiescent water level ($80$\,mm above the channel bed). A third-order polynomial model was used for the PIV camera, while a pinhole model was applied to the FPP camera. The point-LIF camera was calibrated separately using a smaller two-level plate (LaVision 106-10) and a third-order polynomial model.  

The phase--elevation mapping for the profilometry system was established by projecting the fringe pattern onto a flat white plate mounted on a translation stage. Images were recorded at ten vertical positions spanning the quiescent water level of $80$\,mm, with a step size of $5$\,mm. These images were first dewarped, then demodulated to produce phase maps for each height. A second-degree polynomial was then fitted to the phase evolution at each pixel, allowing for the recovery of surface elevation from phase data.  

\begin{figure*}[t]
  \centering
  \includegraphics[width=0.8\textwidth]{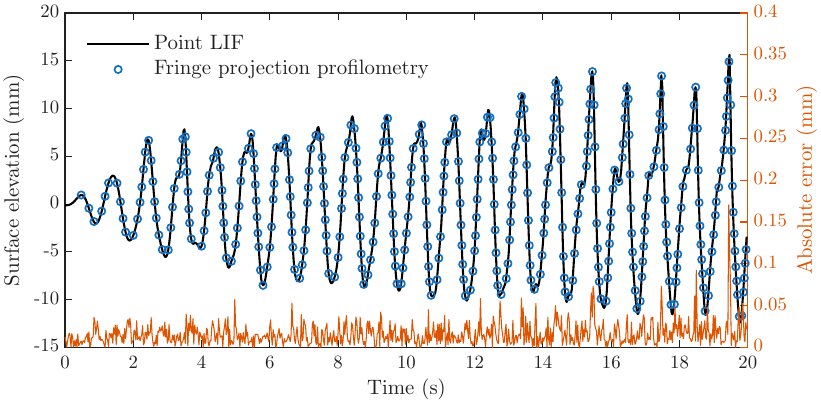}
  \caption{Time series of surface elevation measured by point-LIF and fringe projection profilometry (FPP) at a dye concentration of 12\,mg/L. The absolute difference between the two measurements is shown in orange using the right axis (note the smaller scale).}
  \label{fig:profilometryLIFcomp}
\end{figure*}

We note that, instead of a machined plate, a container with a thin layer of high-concentration dye solution can be used for this calibration. This approach offers two advantages: the liquid surface is self-levelling, and the calibration area can be made as large as required without the manufacturing constraints of a solid plate. A third alternative, simpler than the former two and particularly useful when a translation stage is impractical, utilises the point-LIF system. The water level is raised to the maximum measurement height and the channel is slowly drained while the FPP and point-LIF cameras record at fixed intervals. This yields images spanning the full measurement range and requires no manual intervention beyond opening the drain valve.


\section{Results and Illustrative Applications}\label{sec:Results}
\subsection{Error analysis}\label{sec:ErrorAnalysis}

To assess the accuracy of the FPP method, we compared simultaneous measurements by FPP and point-LIF of water waves generated in initially quiescent water by a wavemaker operating at $1$\,Hz. This procedure ensured that the waveforms were reproducible across runs with different dye concentrations. For each concentration, two sequences of $20$\,s duration were acquired at a frame rate of $40$\,Hz. 

We calculated the characteristic error using amplitude-based binning. This was to ensure that the error analysis represented the full dynamic range of the wave. The absolute difference between FPP and point-LIF data, the error, was partitioned into eight bins according to surface elevation. The error was averaged within each bin, and the mean of these eight bin-averages was defined as the characteristic error corresponding to a particular dye concentration.

Figure~\ref{fig:profilometryLIFcomp} shows a time series of surface elevation measured by the two systems at a dye concentration of $12$\,mg/L (left axis), with the absolute difference plotted on the right vertical axis. The mean absolute error for this time series is \SI{18}{\micro\meter}, with spikes up to \SI{$\approx$ 160}{\micro\meter} towards the end of the run. As mentioned in Sec.~\ref{sec:pointLIF}, the presence of the point-LIF laser spot prevented FPP measurement at the exact same location, necessitating interpolation using data from adjacent points. The observed spikes are the result of interpolation error. While waves propagate essentially unidirectionally along the channel centreline at the start of a run, reflections from the walls eventually create a complex wave field. Sidewall reflections, in particular, induce significant transverse surface undulations, degrading the accuracy of the linear interpolation. We investigated this effect by successively reducing the distance between the two FPP locations and the point-LIF target, calculating the error at each step. This revealed a trade-off: each reduction decreased the interpolation error but simultaneously increased the mean absolute error due to optical interference from the laser spot in the FPP demodulation process. The $4$\,mm separation employed in this study was therefore selected as a compromise between these competing sources of error.

\begin{figure*}[tb]
    \centering
    \begin{subfigure}[t]{0.45\textwidth}
        \centering
        \includegraphics[width=\linewidth]{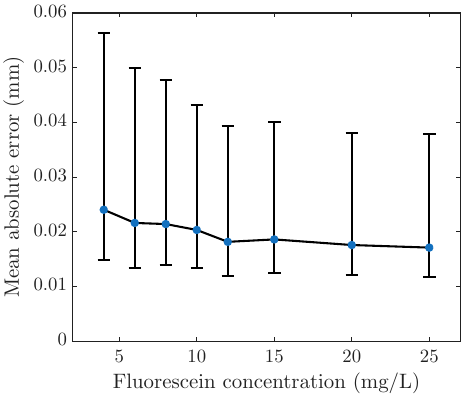}
        \caption{}
        \label{fig:heightErrVsC}
    \end{subfigure}%
  \hfil
    \begin{subfigure}[t]{0.45\textwidth}
        \centering
        \includegraphics[width=\linewidth]{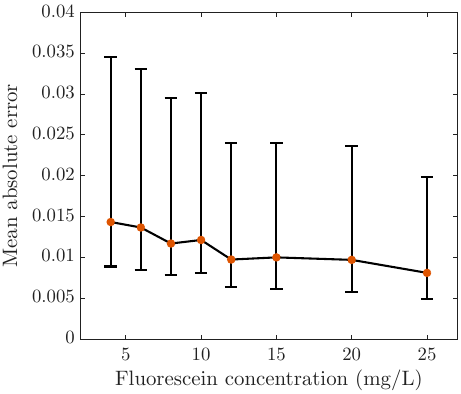}
        \caption{}
        \label{fig:slopeErrVsC}
    \end{subfigure}
    \caption{Variation of mean absolute error of (a) surface elevation and (b) slope with dye concentration. In both figures, the interquartile range (25th to 75th percentiles) of the absolute error is indicated by the error bars.}
    \label{fig:errVsC}
\end{figure*}

Figure~\ref{fig:heightErrVsC} shows how the mean absolute error (MAE) in surface elevation varies with dye concentration. The error bars indicate the interquartile range (25th to 75th percentiles). We observe a modest decrease in MAE with increasing dye concentration until it stabilises at approximately \SI{12}{mg/L}, beyond which the error remains at a plateau of about \SI{18}{\micro\meter}. The 75th percentile of the absolute error decreases more rapidly than the mean, which indicates suppression of large outliers rather than uniform improvement, though this too stabilises near \SI{12}{mg/L}. To interpret this error plateau, we consider the intrinsic uncertainties of the measurement techniques. The noise floor of the FPP system (measured as the standard deviation of a quiescent water surface) is $\sigma_{\mathrm{FPP}} = 8\,\upmu$m. The uncertainty of the point-LIF system is more difficult to quantify, but can be estimated as $\pm 1$\,pixel, or \SI{13}{\micro\meter}. Thus, the lower bound for the MAE is approximately \SI{12}{\micro\meter} (derived from $\sqrt{2/\pi} \sqrt{\sigma_{\mathrm{FPP}}^2 + \sigma_{\mathrm{LIF}}^2}$), close to the \SI{18}{\micro\meter} plateau. The remaining gap between the two is likely attributable to residual calibration errors rather than insufficient dye concentration. Notably, even at the lowest concentration, where light penetration is deepest, the 75th percentile of the elevation error remains below \SI{60}{\micro\meter}. Largely similar trends are observed for the MAE of the slope (Fig.~\ref{fig:slopeErrVsC}), with a modest decrease and a plateau at \SI{12}{mg/L}.

\begin{figure}[tb]
  \centering
  \includegraphics[width=\linewidth]{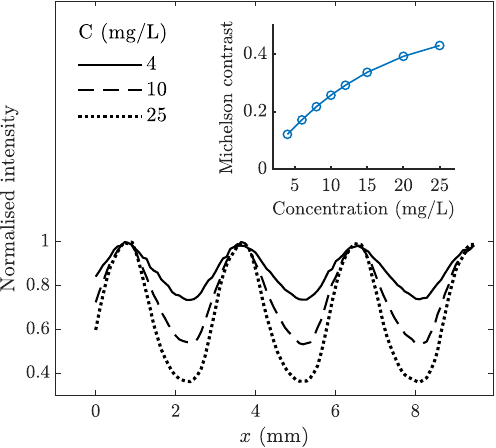}
  \caption{Normalised fringe pattern intensity at three dye concentrations (C = 4, 10, and 25\,mg/L). The inset shows the Michelson contrast as a function of concentration.}
  \label{fig:michelsonContrast}
\end{figure}

To quantify the contrast of the fringe pattern, we utilised the Michelson contrast, defined as $(I_\mathrm{max} - I_\mathrm{min})/(I_\mathrm{max} + I_\mathrm{min})$, where $I_\mathrm{max}$ and $I_\mathrm{min}$ are the local maximum and minimum intensities. Figure~\ref{fig:michelsonContrast} shows the normalised intensity profile for a segment of the fringe pattern at three different concentrations; the inset presents the Michelson contrast as a function of concentration. Despite an almost four-fold increase in contrast from the lowest to the highest concentration, the 75th percentile of the absolute error varies by only \SI{16}{\micro\meter}.

Note, however, that these contrast values were obtained for a quiescent surface. Steep waves or surface roughness can significantly degrade contrast, potentially to the point where the fringe is no longer locally discernible and the demodulation step fails. We therefore recommend a minimum dye concentration of $8$\,mg/L to ensure robust phase recovery.

The addition of fluorescein dye degrades PIV signal quality through two primary mechanisms: (1) excitation of the dye by the laser generates background noise, reducing the SNR, and (2) absorption of laser light by the dye attenuates the laser sheet, further reducing particle visibility. As discussed in Sec.~\ref{sec:opticalFilters}, while the attenuation of laser intensity cannot be avoided, the fluorescence-induced background noise can be mitigated with an ultranarrow bandpass filter.  

\begin{figure}[tb]
  \centering
  \includegraphics[width=\linewidth]{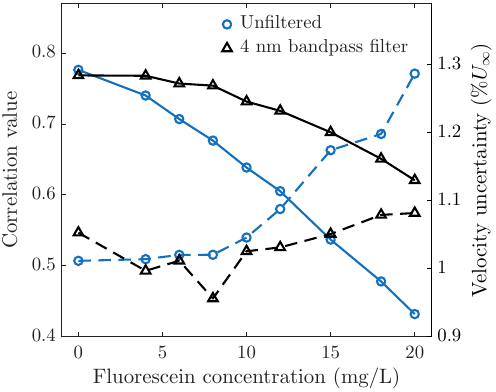}
  \caption{
  Spatially and temporally averaged PIV correlation values (solid lines, left axis) and uncertainty of the velocity magnitude as a percentage of the free stream velocity (dashed lines, right axis) for a uniform flow with and without a bandpass filter on the camera.}
  \label{fig:filterCorrVal}
\end{figure}

To quantify the efficacy of this filter, 500 image pairs were acquired under uniform flow conditions across a range of dye concentrations, both with and without the bandpass filter, as detailed in Sec.~\ref{sec:ExpSetup}. The correlation value was used as a proxy for PIV quality and averaged over time and space to yield a single representative value for each dye concentration, as shown on the left axis in Fig.~\ref{fig:filterCorrVal}. In addition, the uncertainty of the velocity magnitude, calculated from correlation statistics by DaVis 11.0 and shown by \citet{Wieneke2015} to be a good estimate of measurement error, is displayed on the right axis as a percentage of the free stream velocity magnitude, likewise averaged in time and space.

Without fluorescein dye, the mean correlation value is approximately 1\% higher for the unfiltered configuration. This is expected, as the bandpass filter is not perfectly transmissive, resulting in reduced particle image intensity. Once dye is added, however, the filtered case exhibits higher correlation values, with the gap between the two configurations widening as concentration increases. The decline in correlation observed for the filtered case is almost entirely due to the attenuation of the laser sheet by the dye. In contrast, the more rapid degradation in the unfiltered case, and the resulting performance gap, is driven by the background noise generated by fluorescence.  

We note that sufficiently high correlation values can still be obtained without a bandpass filter at lower dye concentrations or for shallow flow depths. Therefore, while an optical bandpass filter consistently improves PIV quality, it may not be strictly necessary to achieve measurements of acceptable accuracy.

A similar divergence between unfiltered and filtered configurations is observed in the velocity uncertainty (dashed lines in Fig.~\ref{fig:filterCorrVal}). The uncertainty remains nearly constant across the full concentration range with a bandpass filter, but increases by 0.3\% without one. While the trend confirms the filter's effect, the increase in error is negligible in this setup, which we attribute to idealised experimental conditions, namely, a shallow optical path (6\,cm of water), large seed particles (\SI{40}{\micro \meter}) combined with high laser power, and uniform flow. Under conditions of greater depth, smaller particles, or higher dye concentration, the benefit of the bandpass filter would be considerably larger.

\subsection{Attenuation length}\label{sec:AttenuationLength}
Subsurface emission of light from the dye reduces the accuracy of surface reconstruction (Sec.~\ref{sec:Semitransparency}). We quantify the subsurface emissions using the attenuation length $\ell$, the distance over which beam intensity falls to $1/e$ of its value at the surface. Absorption follows the Beer-Lambert law, $\log_{10}(I_0/I) = \varepsilon c d$, where $I_0$ and $I$ are light intensities at entry and after path length $d$, $c$ is the molar concentration, and $\varepsilon$ is the molar extinction coefficient. \added{The attenuation length defined here is related to the molar extinction coefficient $\varepsilon$ by $\ell = (\ln (10)\:\varepsilon c)^{-1}$. The molar concentration was calculated using a molar mass of \SI{376.2}{g\,mol$^{-1}$}.}

\begin{figure}[tb]
  \centering
  \includegraphics[width=\linewidth]{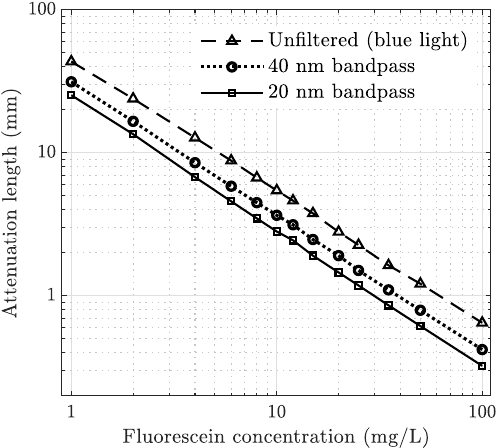}
  \caption{Attenuation length as a function of dye concentration for three projector filter configurations.}
  \label{fig:attenuationLength}
\end{figure}

Since the molar extinction coefficient is tabulated for monochromatic excitation at the peak absorption wavelength and is highly pH sensitive for fluorescein \citep{Mota1991}, we measured the attenuation length directly. A beam from the projector was directed through the left wall of a glass tank of dye solution, propagating horizontally, parallel to and 1\,cm away from the front wall, through which a camera (LaVision sCMOS CLHS) imaged the beam path. We tested three filter configurations: unfiltered, 40\,nm bandpass, and 20\,nm bandpass (see Sec.~\ref{sec:opticalFilters} for details on filters), with results shown in Fig.~\ref{fig:attenuationLength}. The spectral composition of the blue light for the unfiltered case depends on the projector's light source and the dichroic mirrors used for colour separation, details of which were not available from the manufacturer, but should be representative of typical lamp-type commercial projectors.

Relative to the unfiltered configuration, the 40\,nm and 20\,nm bandpass filters reduced the attenuation length by 33\% and 47\%. A linear fit to $\log_{10}{(I_0/I)}$ vs. $d$ yielded \textit{effective} molar extinction coefficients of \SI{31000}, \SI{46000}, and \SI{59000}{L\,mol$^{-1}$\,cm$^{-1}$} at 10\,mg/L for the unfiltered, 40\,nm bandpass and 20\,nm bandpass configurations, respectively. The fit was performed from the entry point to a distance equal to the attenuation length.

At a concentration of just 4\,mg/L, the mean absolute error (\SI{24}{\micro\meter}) is more than two orders of magnitude smaller than the attenuation length (7\,mm). The unexpectedly small error may be explained by the choice of reference image. For each dye concentration, a still-water image was captured and used for demodulation. Such concentration-specific reference images could partly compensate for what would otherwise be a larger systematic error, since surface elevation is measured relative to the reference plane. The geometric centroid of the subsurface emissions is, however, a function of surface topography and viewing angle, meaning that residual errors remain.

\subsection{Illustrative Applications}\label{sec:Applications}
To demonstrate the method, we present two experimental cases involving simultaneous PIV and free-surface profilometry. A further example may be found in \citet{Babiker2026}.

\subsubsection{Flow behind a cylinder}\label{sec:FlowBehindCylinder}

The first case considers the flow behind a cylinder interacting with surface waves. Despite the technological importance of cylinders affected by currents and waves, for example with respect to scour around monopiles, few fundamental studies of the fluid mechanics of the combined system exist. The relatively sparse literature largely considers the flow immediately around the cylinder for purposes of scour prediction, following the seminal work of \citet{sumer1997}, while little direct attention has been given to the lee-side wake. It has been observed, however, that a wave superposed on the current flowing past a cylinder can completely suppress the von K\'{a}rm\'{a}n vortex shedding \citep{gunnoo2016}. More broadly, vortex shedding from bluff bodies has served as a canonical system for studying the coupling between surface deformation and the subsurface velocity field \citep[][]{Dabiri2003,Savelsberg2009, Ng2011}.

A vertical cylinder of diameter $D=7$\,cm was placed in a uniform flow, piercing the free surface. The water contained fluorescein dye at a concentration of $10$\,mg/L. The water surface topography and the underlying flow field in a horizontal plane $2$\,cm beneath the quiescent water level (8\,cm) were captured at a rate of 15\,Hz. The mean flow velocity was $U=0.1$\,m/s, corresponding to a Reynolds number of $Re_D=UD/\nu\approx 7000$. Following a period of steady flow, the wavemaker was activated at $1$\,Hz to generate waves with a wavelength of approximately $83$\,cm, and the interaction between the vortex street and the waves was recorded. 

\begin{figure*}[t]
    \centering
    \begin{subfigure}[t]{0.9\textwidth}
        \centering
        \includegraphics[width=\linewidth]{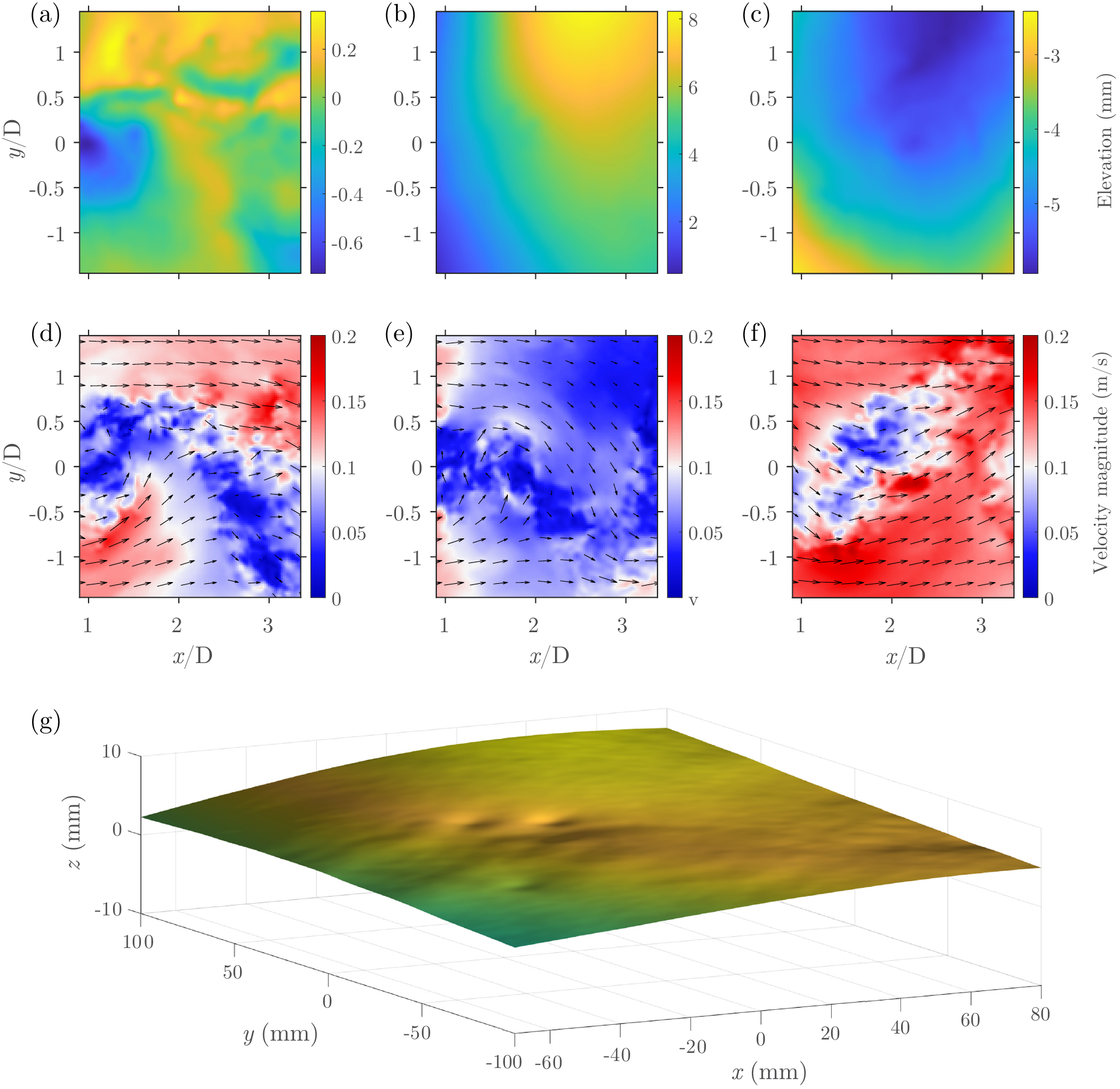}
    \end{subfigure}
    \caption{Application of the technique to flow behind a cylinder at $Re_D \approx 7000$ with waves. (a), (b) and (c) show contours of surface elevation while (d), (e) and (f) show contours of velocity magnitude measured 2~cm beneath the surface (vectors subsampled by a factor of 12 for clarity). Three instants in time are shown: prior to wave arrival (a and d), during the passage of a wave crest (b and e) and trough (c and f).  (g): Visualisation of surface dimples induced by surface-attached vortices advecting over the wave.
    }
    \label{fig:vtxWavesContours}
\end{figure*}

In Fig.~\ref{fig:vtxWavesContours}, contours of surface elevation (top row) and velocity magnitude (middle row) are shown for three instants in time: without waves (left) and at the passing of a wave crest (middle) and trough (right). Before the waves arrive, in panels (a) and (d), the meandering wake of the cylinder is evident, along with flow separation just behind the cylinder. The fall in pressure due to separation of the boundary layer and the ensuing vortex behind the cylinder around $x/D = 1$ generates a depression in the free surface, seen in the contours of surface elevation. After activation of the wavemaker, the waves dominate the motion of the free surface and strongly influence the velocity field. The waves, propagating against the mean current, result in a periodic increase in velocity magnitude beneath wave troughs (panel (f)) and a decrease beneath wave crests (panel (e)). 

Figure~\ref{fig:vtxWavesContours}g shows a perspective view of a wave crest, on top of which three dimples can be seen. These are imprints of surface-attached vortices shed by the cylinder and have a depth of about $0.1$\,mm. The ability to resolve surface perturbations which differ in scale by almost two orders of magnitude is a particular strength of projection methods such as FPP. 

\begin{figure*}[tb]
    \centering
    \begin{subfigure}[t]{0.4\textwidth}
        \centering
        \includegraphics[height=5.6cm]{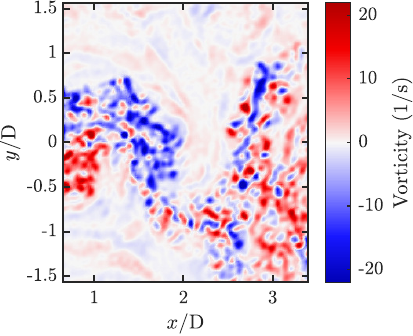}
        \caption{}
        \label{fig:vortBefore}
    \end{subfigure}
    \hfill
    \begin{subfigure}[t]{0.52\textwidth}
        \centering
        \includegraphics[height=5.6cm]{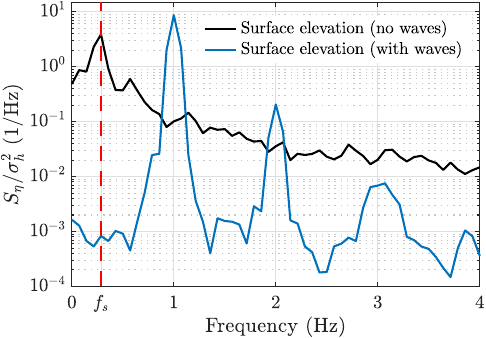}
        \caption{}
        \label{fig:surfSpectrum}
    \end{subfigure}
    \\
    \begin{subfigure}[t]{0.4\textwidth}
        \centering
        \includegraphics[height=5.6cm]{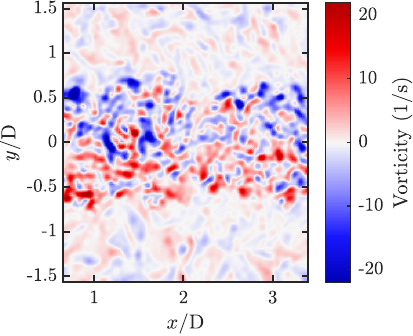}
        \caption{}
        \label{fig:vortAfter}
    \end{subfigure}
    \hfill
    \begin{subfigure}[t]{0.52\textwidth}
        \centering
        \includegraphics[height=5.6cm]{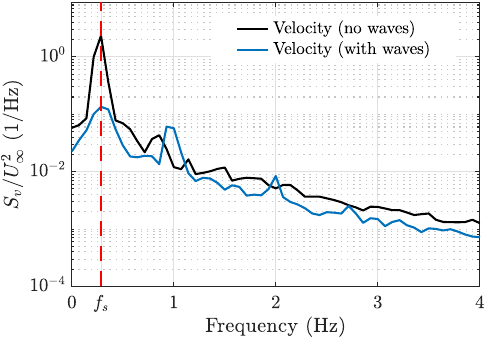}
        \caption{}
        \label{fig:velSpectrum}
    \end{subfigure}
    \caption{Instantaneous subsurface vertical vorticity fields behind a cylinder in turbulent flow in the absence (a) and presence (c) of counterpropagating surface waves. The corresponding power spectral densities are shown for surface elevation (b) and transverse velocity fluctuations (d), with the shedding frequency $f_s$ indicated by the dashed line.}
    \label{fig:vorticitySpectra}
\end{figure*}

Figure~\ref{fig:vorticitySpectra} demonstrates the phenomenon of vortex street suppression with striking clarity. Instantaneous snapshots of the vorticity field before (Fig.~\ref{fig:vortBefore}) and after (Fig.~\ref{fig:vortAfter}) activation of the wavemaker reveal an almost immediate suppression of vortex shedding in the wake; an animation of this process is provided in the supplementary material (Online Resource~1).

To quantify this effect, we computed the power spectral density (PSD) of the surface elevation and the transverse velocity component, as shown in Figs.~\ref{fig:surfSpectrum} and \ref{fig:velSpectrum}, respectively. The PSD was estimated using Welch's method \citep{welch1967}, applying a Hamming window to segments of 210 samples in time with $50$\% overlap, with a total dataset length of 600 samples. The spectra were calculated by spatially averaging the PSDs along lines of constant $y/D$.  For the transverse velocity, the centreline ($y/D = 0$) was used. However, because the shedding signature in the surface elevation is weak along the symmetry line, the elevation spectra were calculated using data along $y/D = 0.5$.

The transverse velocity spectrum (Fig.~\ref{fig:velSpectrum}) exhibits a peak in energy at the shedding frequency, $0.286$\,Hz. This corresponds to a Strouhal number $St = fD/U$ of 0.20, in agreement with the experiments of \citet{Roshko1954}. After activation of the wavemaker, the peak remains but is more than an order of magnitude weaker. Interestingly, a peak at $1$\,Hz and its harmonic at $2$\,Hz are conspicuous in the transverse-velocity spectrum, even though the orbital velocity due to the waves alone has only a very small transverse component from imperfect wave generation and wave scattering from ambient turbulence \citep{Smeltzer2023}. The presence of these peaks indicates coupling between the wave field and the wake dynamics. 

\begin{figure*}[tb]
    \centering
    \begin{subfigure}[t]{0.42\textwidth}
        \centering
        \includegraphics[width=\linewidth]{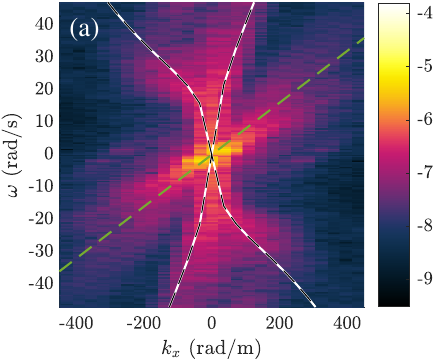}
    \end{subfigure}
    \qquad
    \begin{subfigure}[t]{0.42\textwidth}
        \centering
        \includegraphics[width=\linewidth]{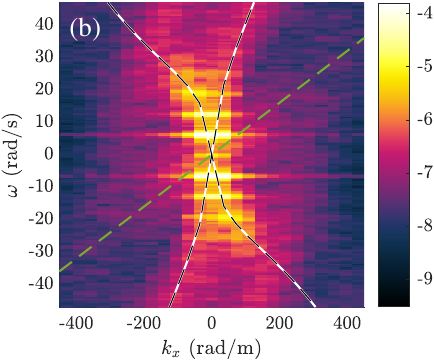}
    \end{subfigure}
    \\
    \vspace{1em}
    \begin{subfigure}[t]{0.42\textwidth}
        \centering
        \includegraphics[width=\linewidth]{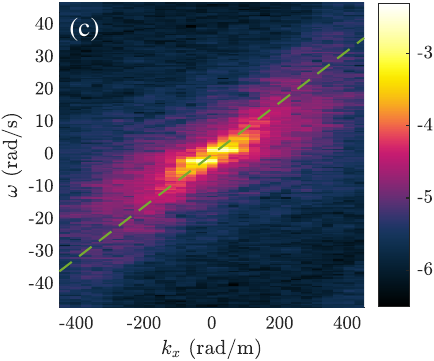}
    \end{subfigure}
    \qquad
    \begin{subfigure}[t]{0.42\textwidth}
        \centering
        \includegraphics[width=\linewidth]{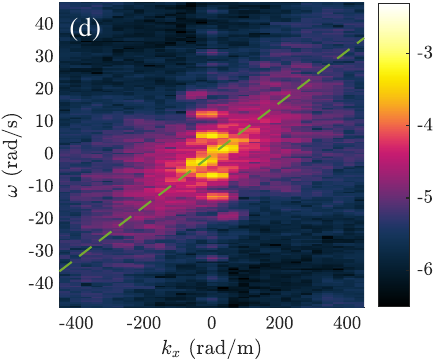}
    \end{subfigure}
    \caption{\added{Wavenumber--frequency spectra $\log_{10} E(k,\omega)$ of the surface elevation (top row) and the transverse velocity component (bottom row). The left column shows pure vortex shedding from a cylinder (a,c), while the right column (b,d) shows vortex shedding coupled with surface waves. The black and white dashed curves show the Doppler-shifted linear dispersion relation, and the dashed green line indicates the convection line $\omega=U_\mathrm{avg} k_x$, where $U_\mathrm{avg} = 0.08$\,m/s.}}
    \label{fig:wavenumberFreqSpectra}
\end{figure*}

The surface elevation spectrum (Fig.~\ref{fig:surfSpectrum}) shows a peak at the shedding frequency (0.286\,Hz) prior to wavemaker activation. After wave generation begins, the power spectrum is dominated by the waves, seen as peaks at the frequency of the wavemaker ($1$\,Hz) and its harmonics. This is expected, given that the wave amplitude is almost two orders of magnitude larger than the oscillations resulting from vortex shedding. Notably, however, there is no longer an observable peak at the shedding frequency in the surface elevation spectrum.

\added{
Space- and time-resolved measurements allow analysis beyond the frequency spectra of Fig.~\ref{fig:vorticitySpectra}. Figure~\ref{fig:wavenumberFreqSpectra} shows wavenumber--frequency spectra at $k_y=0$ (i.e.,\ along the streamwise direction) for surface elevation (top row) and transverse velocity (bottom row). The green dashed line indicates the convection line $\omega=U_\mathrm{avg} k_x$ at $U_\mathrm{avg}=0.08$\,m/s \citep[see, e.g.][]{Bullee2024}, where $U_\mathrm{avg}$ is the average streamwise velocity over the measurement domain. This is smaller than the free-stream velocity of 0.1\,m/s due to the velocity deficit in the cylinder wake. The Doppler-shifted linear dispersion relation is overlaid as black and white dashed curves. The surface elevation spectra were computed over $0.5 < y/D < 1$, whereas the velocity spectra were computed over the full measurement domain. Before wave generation, the surface spectrum, panel (a), shows energy distributed along both the dispersion relation (due to ambient waves) and the convection line, with distinct peaks at the shedding frequency. Velocity variations in the wake of the cylinder smear spectral energy around the convection line, as seen in both surface and velocity spectra. The transverse velocity spectrum, panel (c), also exhibits shedding peaks, with a corresponding streamwise spatial period of approximately 20\,cm. After wave generation (right column), the surface spectrum, panel (b), is dominated by the wave frequency and its harmonics. The footprint of the waves is also seen in the velocity spectrum, panel (d). However, unlike the surface spectrum, the shedding frequency remains detectable, though at considerably diminished amplitude, consistent with Fig.~\ref{fig:velSpectrum}.
}

\citet{gunnoo2016} remark that suppression of the von  K\'{a}rm\'{a}n vortex shedding can occur with sufficiently large wave amplitudes, but the physical process behind this suppression remains unclear. In the present work, the interaction between waves and vortex shedding is presented primarily to demonstrate the capabilities of the measurement technique. Indications are, however, that this interaction could be fertile ground for future studies.


\subsubsection{Droplet impact}\label{sec:Droplet}

\begin{figure*}[tb!]
    \centering
    \begin{subfigure}[t]{0.86\linewidth}
        \centering
        \includegraphics[width=\linewidth]{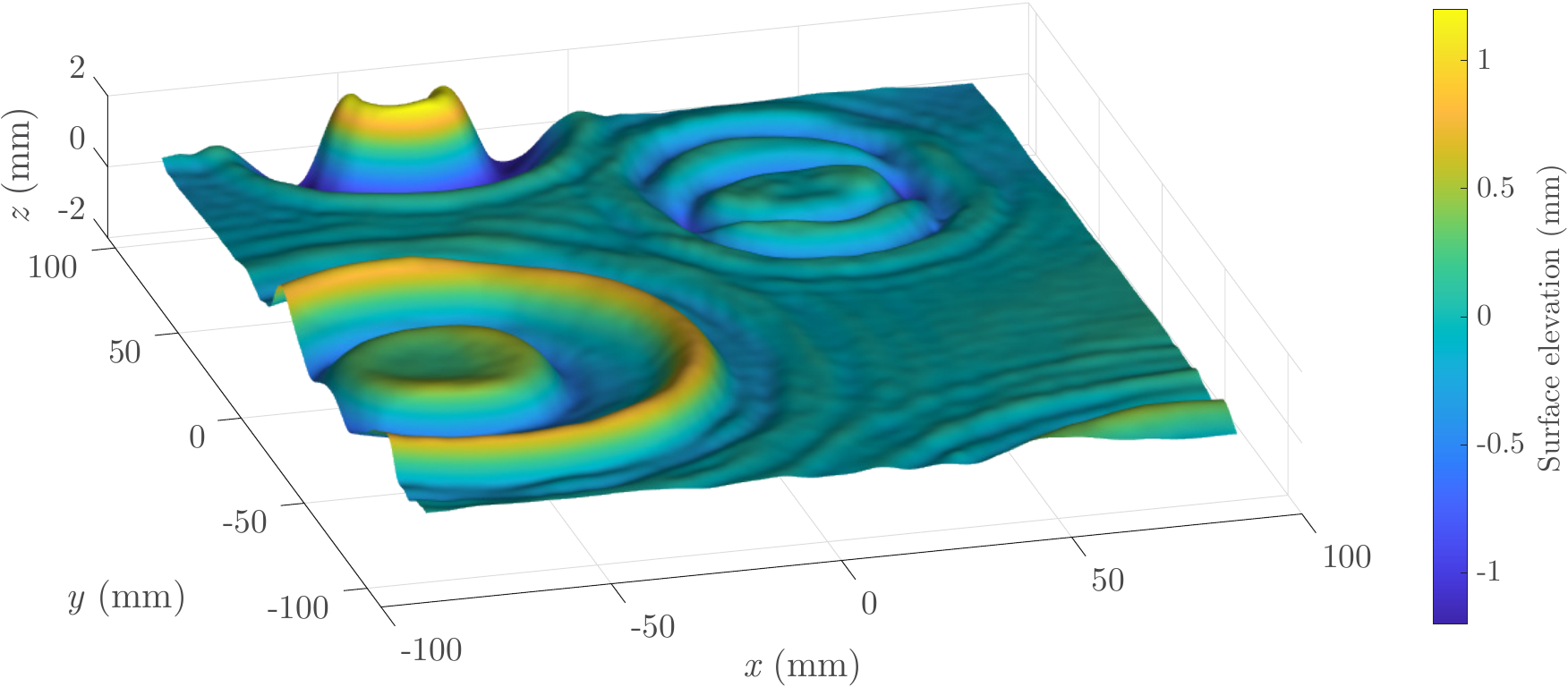}
        \caption{}
        \label{fig:dropletSurf3D}
    \end{subfigure}
    \\
    \begin{subfigure}[t]{0.4\linewidth}
        \centering
        \includegraphics[height=5.8cm]{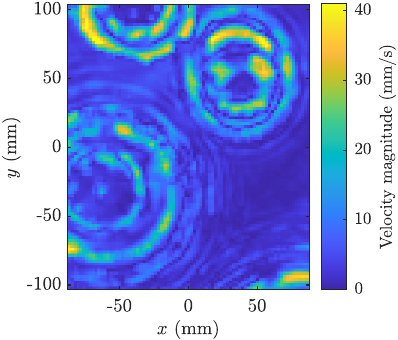}
        \caption{}
        \label{fig:dropletVelocity}
    \end{subfigure}
    \hspace{3.9em}
   \begin{subfigure}[t]{0.4\linewidth}
        \centering
        \includegraphics[height=5.8cm]{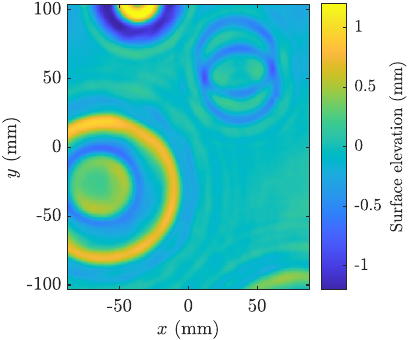}
        \caption{}
        \label{fig:dropletSurf2D}
    \end{subfigure}
    \vspace{1em}
    \caption{Application of the technique to droplet impacts on an initially quiescent surface. We let droplets fall randomly, and the frame shown is an arbitrarily chosen snapshot. (a) Three-dimensional reconstruction of the free surface. (b) Contours of horizontal velocity magnitude measured 1\,cm below the surface. (c) Contours of surface elevation relative to the quiescent water level.}
    \label{fig:dropletImpact}
\end{figure*}

\begin{figure}[tb]
    \centering
    \includegraphics[width=\linewidth]{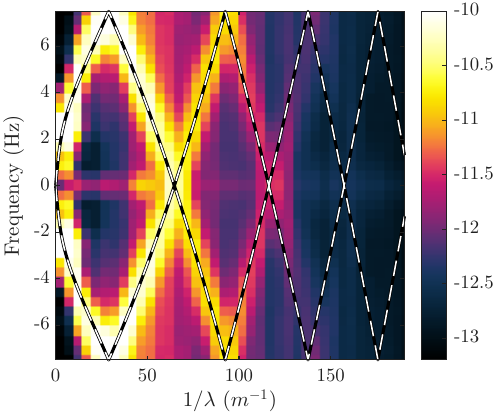}
    \caption{Wavenumber--frequency spectrum $\log_{10} E(k,\omega)$ for waves resulting from random droplet impacts. The curves show the linear dispersion relation, artificially aliased to facilitate comparison with experimental data.}
    \label{fig:dropletSpectra}
\end{figure}

As the second demonstration case, we measure the surface deformation and subsurface velocity field during the impact of falling droplets on an initially quiescent water surface. The impact of a falling droplet on a free surface has previously been used as a benchmark for simultaneous surface and subsurface measurements \citep{Steinmann2021}. The water depth was 8\,cm, the subsurface measurement plane was located 1\,cm beneath the free surface and images were recorded at 15\,Hz. 

Figure \ref{fig:dropletSurf3D} shows a 3D reconstruction of the surface at a single instant. The interference pattern of two ring waves is apparent on the top right of the plot. Note that surface amplitudes are highly exaggerated by the different axis scaling, for visibility. Capillary waves of amplitude as low as \SI{50}{\micro\meter} can be seen at the same time as the much larger and steeper ring waves on the top and bottom left. In Fig.~\ref{fig:dropletVelocity}, contours of velocity magnitude are shown for the same instant in time, while Fig.~\ref{fig:dropletSurf2D} shows contours of surface elevation. While the subsurface velocity measurements exhibit some noise due to the highly three-dimensional nature of the impact flow, the spatial correspondence between the surface deformation and the flow features remains clear.

A wavenumber--frequency spectrum $E(k, \omega)$ of the surface, overlaid with the linear dispersion relation, allows us to estimate the spatial resolution limit of the profilometry. The spectrum in Fig.~\ref{fig:dropletSpectra} was computed using a recording containing 600 snapshots (40 seconds) of surface elevation data. Due to the 15\,Hz acquisition rate (limited by the laser repetition rate), the experimental spectrum is temporally aliased. To enable a direct comparison, we folded the theoretical dispersion curve to match this aliasing, resulting in the sawtooth-like appearance of the curve. The experimental energy distribution shows excellent agreement with the theoretical curve up to a wavenumber of $1/\lambda = 150\,\mathrm{m}^{-1}$, corresponding to a wavelength of 6.7\,mm. While it is unclear whether this limit arises from the 2.7\,mm fringe spacing or the spectral noise floor, Fig.~\ref{fig:dropletSpectra} confirms that capillary waves are accurately resolved down to wavelengths of $6.7$\,mm.

\section{Challenges and practical recommendations}\label{sec:Challenges}
This section outlines specific technical challenges encountered during the experimental campaign and offers practical recommendations for implementing this technique.

\begin{itemize}
    \item Projector filter bandwidth: The attenuation length of light was less critical to measurement accuracy than initially anticipated. As a result, the \SI{20}{nm} bandwidth filter proved unnecessarily restrictive. The 40\,nm filter transmits more light, yielding a brighter projected pattern and a higher signal-to-noise ratio. Despite the accompanying reduction in fringe contrast from increased subsurface emissions, this brightness advantage dominates, and the 40\,nm filter yielded a slightly lower MAE. When the projector input was adjusted to equalise the output brightness of both configurations, thereby removing the SNR advantage of the 40\,nm filter, the 20\,nm filter provided only a marginal reduction in MAE (on the order of a few micrometres).

    \item Projector type: Modern high-brightness commercial video projectors are typically laser based, using either three separate red, green and blue lasers or a single blue laser with a phosphor wheel, and in either case emit comparatively little light in fluorescein's primary absorption range (cyan, approximately 480--500\,nm). One should bear in mind that the relevant quantity is the spectral overlap between emitted light and fluorescein's absorption spectrum, not the total luminous flux (lumens). Lamp-type projectors (metal-halide or high-pressure mercury vapour, often branded ``UHP'' or ``UHE'') on the other hand, emit broad-spectrum white light that overlaps with the primary absorption band of fluorescein.
    
    \item Light source intensity: While commercial projectors offer convenience, their limited brightness prevents the use of very short exposure times. In experiments on wind-driven waves (not detailed in this manuscript), we successfully used our set-up as described, with an exposure time of \SI{500}{\micro s} and an acquisition frequency of \SI{2}{kHz}. However, for applications requiring significantly shorter exposure times, a high-power laser-based light source (for instance similar to that of \citet{Roth2020}) could be necessary to achieve sufficient brightness. This might require a number of modifications to other parts of the set-up, beyond our present scope.

    \item Projector placement: Care must be taken to orient the projector so that the exhaust fan does not direct hot air across the optical path of the profilometry camera. Turbulent hot air induces refractive index fluctuations, resulting in image jitter that can easily be mistaken for surface movement.
    
    \item Resolution limit: Drawing from the experiments presented in this work, as well as additional experiments, the effective spatial resolution limit is approximately twice the fringe wavelength. To incorporate a margin of safety, we recommend selecting a fringe wavelength no larger than 0.4 times the smallest length scale of interest $L_\mathrm{min}$, i.e., $\lambda_{f} \leq 0.4 L_\mathrm{min}$.

    \item 
    Surface contaminants: Bubbles or flecks of material floating on the surface appear as bright regions in the profilometry images, causing large local perturbations in the phase field. Contaminants should therefore be minimised where possible; however, some contamination is often unavoidable, particularly in large flumes. To address this, our processing code detects anomalously large spatial gradients in the phase field and logs their locations. These flagged regions can then be masked out of the resulting surface elevation fields.
\end{itemize}

\section{Conclusions}

We have demonstrated a new method for simultaneous measurement of a moving free water surface and the subsurface velocity field, using a combination of Fringe Projection Profilometry (FPP) and particle image velocimetry (PIV). The method involves dyeing the water with a fluorescent dye (fluorescein) which effectively makes the water opaque to some wavelengths of light and transparent to others. Fluorescein strongly absorbs blue wavelengths and emits in the green.

A cyan sine-wave pattern is projected onto the surface by a video projector; the dye's strong absorption of this light allows the surface elevation to be deduced from the displacement of the observed pattern. A PIV light sheet is created with a green laser of wavelength $532$\,nm, where fluorescein has weak, but nonzero, absorption.

A combination of optical filters allows for clean images for both PIV and profilometry. A long-pass filter was applied to the FPP camera to reject specular reflections (which are cyan) and retain only the fluorescence (which is green). Some of the green laser light was absorbed and reemitted, causing noise in the PIV images; we effectively removed this with a narrowband filter centred at the laser's wavelength. 

The surface elevation measurements were validated against single-point Laser-Induced Fluorescence (LIF) measurements. We found a mean absolute error (MAE) of \SI{17}{\micro\meter} at the highest fluorescein concentration, $25$\,mg/L. As expected, errors increase at lower concentrations as the attenuation length of the projected light increases. However, they do so surprisingly slowly; even at a concentration of $4$\,mg/L the 75th percentile of the error remains below \SI{60}{\micro\meter} and the MAE is \SI{24}{\micro\meter}, sufficient for many purposes. Beyond approximately $12$\,mg/L, where the MAE is \SI{18}{\micro\meter}, increasing the concentration of fluorescein did not result in an appreciable improvement in accuracy. In terms of PIV performance, the correlation values for our set-up were essentially unchanged up to concentrations of $8$\,mg/L. Beyond this point, correlation values declined but remained robust ($>\!0.6$ at 20\,mg/L) due to the effective noise suppression of the narrowband filter.

The capabilities of the method were demonstrated through two experimental cases. First, we measured flow past a vertical cylinder at a Reynolds number of $Re_D = UD/\nu\approx 7000$. We observed that the addition of surface waves could suppress the von K\'arm\'an vortex street behind the cylinder. Notably, the system was able to resolve small dimples from surface-attached `bathtub' vortices, 0.1\,mm deep atop waves of centimetre amplitude, demonstrating the ability of FPP to measure accurately across disparate scales. As a second example, we captured the simultaneous surface ring waves and subsurface velocity fields produced by droplet impacts. The technique therefore proves to be both accurate and practical, offering a versatile solution for a wide range of fluid dynamics applications.


\subsection*{Acknowledgements}
The members of the Ellingsen/Hearst research groups contributed to discussions throughout, in particular Dr Stefan Weichert in the early phase, and later Drs Am\'{e}lie Ferran, Leon Li and Olav R\o mcke. 

\subsection*{Author contribution statement}

The authors confirm contribution to the paper as follows. A.~Semati developed the combined FPP and PIV methodology, performed the experiments and is the main author of the manuscript. A.~Shankaran contributed to the development and implementation of the experimental setup. B.~K.~Smeltzer had the original idea behind the method and carried out preliminary tests. E.~\AE s\o y implemented the initial FPP method and developed the data analysis framework. R.~J.~Hearst advised the PIV and LIF work, including testing and benchmarking. S.~\AA.~Ellingsen supervised the project, contributed to discussions throughout, and co-wrote the manuscript.

\noindent\textit{All authors reviewed the results and approved the final version of the manuscript.}

\subsection*{Funding}
The work is co-funded by the Research Council of Norway (STF, \emph{iMOD}, 325114) and the European Union (ERC CoG, \emph{WaTurSheD}, 101045299 and ERC StG \emph{GLITR}, 101041000). Views and opinions expressed are, however, those of the authors only and do not necessarily reflect those of the European Union or the European Research Council. Neither the European Union nor the granting authority can be held responsible for them.

{
\small
\subsection*{Data availability}
The data supporting the findings of this study are available for download from the Norwegian national data repository DataverseNO at \href{https://doi.org/10.18710/MWEHEM}{doi.org/10.18710/MWEHEM}.

\subsection*{Code availability}
The code used to generate the surface elevation data presented in this study is publicly available on GitHub (\href{https://github.com/asemati/FTProfilometry}{https://github.com/asemati/FTProfilometry}).
}




\end{document}